\documentclass[12pt]{article}
\usepackage{amsmath}
\usepackage{graphics}
\usepackage{epstopdf, color}

\begin{document}
\title{Ladder operators depending on all variables for a charged particle moving in a two-dimensional uniform magnetic field}
\author{Shishan Dong$^1$, B. J. Falaye$^{2}$\thanks{E-mail address: fbjames11@physicist.net}, A. E. Guerrero M.$^{3}$
and
Shi-Hai Dong$^{3}$\thanks{E-mail address: dongsh2@yahoo.com}\\
{\footnotesize $^{1}$Information and Engineering College, Dalian University, Dalian 116622, P. R. China}\\
{\footnotesize $^{2}$ESFM, Instituto Polit\'ecnico Nacional, Edificio 9, Unidad Profesional ALM, Mexico D. F. 07738, Mexico }\\
{\footnotesize $^{3}$CIDETEC, Instituto Polit\'ecnico Nacional,
UPALM, Mexico D. F. 07700, Mexico}}

\date{}
\maketitle

\begin{abstract}
The ladder operators for one dimensional quantum harmonic oscillator were constructed by Schr\"{o}dinger in 1940s. We extend this method to a two dimensional uniform magnetic field and establish
the ladder operators which depend on all spatial variables of quantum system. The Hamiltonian of quantum system can also be written by the velocity of the particle.

\end{abstract}
\maketitle

\section{Introduction}
The algebraic method has become the subject of interest in various fields\cite{SH1, SH2, SH3, SH4, SH5, SH6, ques, naila, ricardo, jose, fakh}. A main approach to algebraic method is related with the factorization method \cite{SH7}, in which we have obtained the ladder operators for many solvable potentials including the Morse potential, the P\"{o}schl-Teller potential, the square-well potential, etc. We refer the readers to ref. \cite{SH7} for an elaborate background information about ladder operators. Moreover, it should be stressed that the ladder operators for these quantum systems mentioned above depend on only one variable. That is to say, they are only concerned with one-dimensional problem even though some quantum systems are treated in two and/or three dimensions but the ladder operators are relevant for one variable either radial part or other angular part.

Recently, we have generalized this method to the ladder operators that depend on all spatial variables. The typical examples are the circular and spherical wells directly from the normalized wave functions \cite{SH8, SH9, SH10}.
Essentially, these ladder operators constructed in Refs. \cite{SH8, SH9, SH10}  are called shift operators. The main difference between them lies in the fact that the ladder operators are concerned with the different energy spectra within the same potential, while the shift operators refer to a group of potentials connected with the same energy. With the same spirit, our aim is to study the ladder operators depending on all variables $\rho$ and $\varphi$ realized for a charged particle moving in a two-dimensional uniform magnetic field. This arises from the recent interest in the lower-dimensional field theory and condensed matter physics \cite{ernst1, somo} and also from our previous study on the ladder operators for this problem, which essentially is one dimensional case \cite{SH11}. It is worth emphasizing that we are only interested in the case where a uniform magnetic field $B$ and the electric field $V(\rho)=0$ are considered as before \cite{SH11, SH12, LD}.

This work is organized as follows. In Sect. 2 we present the exact solutions of a charged particle in a uniform magnetic field in two dimensions. In Sect. 3 we present general formalism to establish the ladder operators. Some concluding remarks are given in Sect. 4.

\section{Exact solutions}
We begin by considering potential model $V(\rho)$ subjected to the electromagnetic field \cite{SH12}
\begin{equation}\label{E1}
\frac{1}{2\mu}\left(\vec{p}-q\vec{A}\right)^2\psi=[E-V(\rho
)]\psi,
\end{equation}
where $\vec{p}=-i\hbar \vec {\nabla}$ and $q$ denotes charge. The $\vec{A}$ and $V(\rho)$ denote the vector potential and electric field, respectively. In this work, we are only interested in the case with a uniform magnetic field $B$ and the electric field $V(\rho)=0$ \cite{SH12, LD}. As shown by Landau and Lifshitz \cite{LD}, we take the vector potential of the form $A_{\varphi}=B\rho/2$, $A_{\rho}=0$. Expanding equation (\ref{E1})
and using the property $\vec{\nabla}\cdot(\vec{A}\psi)=\vec{A}\cdot\vec{\nabla}\psi+\psi\vec{\nabla}\cdot\vec{A}$ and $\vec{\nabla}\cdot\vec{A}=0$, Eq. (\ref{E1}) becomes
\begin{equation}\label{E4}
-\hbar^2{\nabla}^2\psi+2i\hbar\, q\vec{A}\cdot\vec{\nabla}\psi+q^2\vec{A}\cdot\vec{A}\psi=2\mu\, E\psi.
\end{equation}
By taking $\psi(\vec{\rho})=(1/\sqrt{2\pi})\rho^{-1/2}\mathcal{H}(\rho)e^{im\varphi}$ ($m=0, \pm 1, \pm 2, \ldots$), we obtain
\begin{equation}\label{E6}
\displaystyle -\hbar^2\mathcal{H}''(\rho)-\frac{\hbar^2\mathcal{H}(\rho)}{4\rho^2}+\frac{m^2\hbar^2\mathcal{H}(\rho)}{\rho^2}-\frac{2\hbar m\, q}{\rho}\left(\frac{B\rho}{2}\right)\mathcal{H}(\rho)+q^2\vec{A}\cdot\vec{A}\mathcal{H}(\rho)=2\mu\, E\mathcal{H}(\rho).
\end{equation}
For electron, $q=-e$ and using atomic units $e=\hbar=\mu=1$,  we obtain a simpler expression
\begin{equation}\label{E8}
\mathcal{H}''(\rho)+\left(\beta-\frac{m^2-\frac{1}{4}}{\rho^2}-2\sigma\, m-\sigma^2\, \rho^2\right)\mathcal{H}(\rho)=0, ~~~\beta=2E, ~~~\sigma=B/2,
\end{equation}which can be further written as
\begin{equation}\label{E9}
\zeta\mathcal{H}''(\zeta)+\frac{1}{2}H'(\zeta)+\left(\frac{\tau}{4\sigma}-\frac{m^2-1/4}{4\zeta}-\frac{\zeta}{4}\right) \mathcal{H}(\zeta)=0, ~~~~\zeta=\sigma\, \rho^2, \end{equation}
where $\tau=\beta-2\sigma\, m$. By taking the wave function $H(\zeta)=\zeta^{1/4+|m|/2}e^{-\zeta/2}G(\zeta)$, we have
\begin{equation}\label{E10}
\zeta G''(\zeta)+\left(1+|m|-\zeta\right) G'(\zeta)+G(\zeta) \left(\lambda -\frac{|m|+1}{2}\right)=0, ~~~~\lambda=\frac{\tau}{4\sigma}=\frac{\beta}{4\sigma}-\frac{m}{2},
\end{equation}
whose solution is given by {$\, _1F_1\left[-\left(\lambda -{(|m|+1)}/{2}\right);|m|+1;\zeta\right]$. From quantum condition $\lambda-(|m|+1)/2=n_{\rho}$, we have
\begin{equation}\label{E11}
E_{n_{\rho}m}=B\left(n_{\rho}+\frac{m}{2}+\frac{|m|+1}{2}\right), ~~~~n_{\rho}=0, 1, 2, \ldots.
\end{equation}
In the case of negative $m$, we have $E_{n_{\rho}}=B(n_{\rho}+1/2)$. For positive $m$, however, we have $E_{n_{\rho}m}=B(n_{\rho}+m+1/2)$. The incorporating them allows us to have $E_{n}=B(n+1/2), n=0, 1, 2, \ldots$. However,
the whole wave functions are calculated as
\begin{equation}\label{E16}
\psi_{n_{\rho}m}(\vec{\rho})=\frac{1}{\sqrt{2\pi}}\sqrt{\frac{2\sigma n_{\rho}!}{(n_{\rho}+|m|)!}}\, e^{-\zeta/2} \zeta^{|m|/2}\, L_{n}^{|m|}(\zeta)e^{im\varphi}.
\end{equation}

Let us study this problem from another point of view. If defining the velocity of the particle as $\hat{v}=(\vec{p}+\vec{A})/\mu$, then the Hamiltonian can be expressed as $H=\mu \hat{v}^2/2$, which is just the kinetic energy of the particle.
Using the following commutation rules
\begin{equation}\label{E32}
[\hat{v}_{x}, \hat{v}_{y}]=-i\left(\frac{\partial A_{y}}{\partial x}-\frac{\partial A_{x}}{\partial y} \right)=-i\, B, ~~~ [\hat{v}_{y}, \hat{v}_{z}]=0, ~~~[\hat{v}_{z}, \hat{v}_{x}]=0,
\end{equation}
the Hamiltonian is expressed by
\begin{equation}\label{E33}
H=\frac{1}{2}(\hat{v}_{x}^2+\hat{v}_{y}^2), ~~~\mu=\hbar=1.
\end{equation} Introducing $\hat{v}_{x}=\sqrt{B}\hat{Q}, \hat{v}_{y}=\sqrt{B}\hat{P}$ and using $[\hat{Q}, \hat{P}]=i$, the Hamiltonian $H=(B/2)(\hat{Q}^2+\hat{P}^2)$. Its energy coincides with the above result.

\section{Ladder operator}

We now address the problem of finding the ladder operators for the wave functions (\ref{E16}) by the factorization method. As shown in Ref. \cite{SH7}, the ladder operators can be constructed directly from the wave functions using only one variable, while in the present study the ladder operators depend on all variables, $\rho$ and $\varphi$. Our aim is to establish the connection between the wave functions $\psi_{nm}(\vec{\rho})$ and $\psi_{(n\pm 1)\, (m\pm 1)}(\vec{\rho})$ through the ladder operators, i. e. $\psi_{(n\pm 1)\, (m\pm 1)}(\vec{\rho})=\mathcal{L}^{\pm}\psi_{n\, m}(\vec{\rho})$, but this is more complicated in comparison with the famous harmonic oscillator case $\psi_{n}$ and $\psi_{n\pm 1}$ worked out first by Schr\"{o}dinger in 1940. For simplicity, we denote quantum number $n_{\rho}$ by $n$.

To this end, we start by acting the following operator on the wave functions (\ref{E16})
\begin{equation}\label{E17}
{\displaystyle\left(\frac{\partial}{\partial x}\pm
i\frac{\partial}{\partial y}\right)\psi_{n_{\rho}m}(\rho, \varphi)= \displaystyle e^{\pm
i\varphi}\left[\frac{\partial}{\partial\rho}\pm \frac{i}{\rho}\frac{\partial}{\partial\varphi}\right]\psi_{n_{\rho}m}(\rho, \varphi), }
\end{equation}
where we have used the following relations
\begin{equation}\label{E18}
x=\rho\cos(\varphi), ~~~y=\rho\sin(\varphi), ~~~~\tan(\varphi)=\frac{y}{x}, ~~~~\zeta=\sigma \rho^2, ~~~\rho=\frac{\sqrt{\zeta}}{\sqrt{\sigma}}.
\end{equation}

First, let us establish the relation between the wave function $\psi_{n\, m}(\rho, \varphi)$ and $\psi_{(n+1)\, (m+1)}(\rho, \varphi)$. To {achieve this}, we have to consider the following recurrence relations {of} the associated Laguerre polynomials \cite{SH13}
\begin{eqnarray}
\label{E19}
&&\frac{d}{dx}L_{n}^{\alpha}=\frac{n\, L_{n}^{\alpha}(x)-(n+\alpha)L_{n-1}^{\alpha}(x)}{x}, \nonumber\\
&&x\, L_{n}^{\alpha+1}(x)=(n+\alpha)L_{n-1}^{\alpha}(x)-(n+\alpha+1-x)L_{n}^{\alpha}(x), \\
&&(n+1)L_{n+1}^{\alpha}(x)-(2n+\alpha+1-x)L_{n}^{\alpha}(x)+(n+\alpha)L_{n-1}^{\alpha}(x)=0, \nonumber
\end{eqnarray}
from which we have
\begin{equation}
\label{E20}
\frac{d}{dx}L_{n}^{\alpha}(x)=\frac{(2+\alpha+2n-x)}{(1+n-x)}L_{n}^{\alpha}(x)-\frac{n+1}{n+1-x}L_{n+1}^{\alpha+1}(x).
\end{equation}

Using above formula (\ref{E20}), we are able to obtain the following relation
\begin{equation}\label{E21}
\begin{array}{l}
-\displaystyle\frac{(n+1-\zeta)}{2\sqrt{\sigma}}\Big[\frac{\partial}{\partial \rho}+\frac{i}{\rho}\frac{\partial}{\partial \varphi}-2\sqrt{\sigma\zeta}\left(\frac{1}{2}-\frac{n+|m|/2+1}{\zeta}+\frac{(n+1)(n+|m|+1)}{\zeta(n+1-\zeta)}\right)\\[3mm]
+\displaystyle\frac{m\sqrt{\sigma}}{\sqrt{\zeta}}\Big]e^{i\varphi}\psi_{nm}
=\sqrt{(n+1)(n+|m|+1)(n+|m|+2)}\psi_{(n+1)(m+1)},
\end{array}
\end{equation}
where we used $n$ to replace $n_{\rho}$ for simplicity. Furthermore, considering the relation (\ref{E18}), we can simplify the above expression as
\begin{equation}\label{E22}
\begin{array}{l}
\displaystyle(n+1-\zeta)\sqrt{\zeta}\Big[-\frac{\partial}{\partial \zeta}+\left(\frac{1}{2}-\frac{n+|m|/2+1}{\zeta}+\frac{(n+1)(n+|m|+1)}{\zeta(n+1-\zeta)}\right)\Big]e^{i\varphi}\psi_{nm}\\[3mm]
=\sqrt{(n+1)(n+|m|+1)(n+|m|+2)}\psi_{(n+1)(m+1)},
\end{array}
\end{equation}
from which we define
\begin{equation}\label{E23}
{\mathcal{L}^{+}}=\displaystyle(n+1-\zeta)\sqrt{\zeta}\Big[-\frac{\partial}{\partial \zeta}+\left(\frac{1}{2}-\frac{n+|m|/2+1}{\zeta}+\frac{(n+1)(n+|m|+1)}{\zeta(n+1-\zeta)}\right)\Big]e^{i\varphi}{, }
\end{equation}
with the property
\begin{equation}\label{E24}
{\mathcal{L}^{+}}\psi_{nm}=\sqrt{(n+1)(n+|m|+1)(n+|m|+2)}\psi_{(n+1)(m+1)}.
\end{equation}

On the other hand, to find the relation between $\psi_{n_{\rho}m}(\rho, \varphi)$ and $\psi_{(n_{\rho}-1)\, (m-1)}(\rho, \varphi)$ we have to consider the following relations except for the first formula used in (\ref{E19})
\begin{eqnarray}
\label{E25}
&&L_{n}^{\alpha-1}(x)=L_{n}^{\alpha}(x)-L_{n-1}^{\alpha}(x)\ \ \mbox{and} \\
&&n\, L_{n}^{\alpha}(x)=(2n+\alpha-1-x)L_{n-1}^{\alpha}(x)-(n+\alpha-1)L_{n-2}^{\alpha}(x){. }
\end{eqnarray}
{Consequently, } we have
\begin{equation}\label{E26}
\frac{d}{dx}L_{n}^{\alpha}(x)=\frac{n+\alpha-1}{n-x}\left[\frac{n+\alpha}{x}L_{n-1}^{\alpha-1}(x)-\frac{n(\alpha+x)}{x(n+\alpha-1)}L_{n}^{\alpha}(x)\right].
\end{equation}

In a similar way, we have the following relation
\begin{equation}\label{E27}
\begin{array}{l}
\displaystyle\frac{n-\zeta}{2\sqrt{\sigma}}\Big[\frac{\partial}{\partial \rho}-\frac{i}{\rho}\frac{\partial}{\partial \varphi}-2\sqrt{\sigma\zeta}\left(-\frac{1}{2}+\frac{|m|}{2\zeta}
-\frac{n(|m|+\zeta)}{\zeta(n-\zeta)}\right)\\[3mm]
-\displaystyle\frac{|m|\sqrt{\sigma}}{\sqrt{\zeta}}\Big]e^{-i\varphi}\psi_{nm}
=\sqrt{n(n+|m|)(n+|m|-1)}\psi_{(n-1)(m-1)},
\end{array}
\end{equation}
from which one is able to obtain a simple form
\begin{equation}\label{E28}
\displaystyle(n-\zeta)\sqrt{\zeta}\Big[\frac{\partial}{\partial \zeta}-\left(-\frac{1}{2}+\frac{|m|}{2\zeta}
-\frac{n(|m|+\zeta)}{\zeta(n-\zeta)}\right)\Big]e^{-i\varphi}\psi_{nm}=\sqrt{n(n+|m|)(n+|m|-1)}\psi_{(n-1)(m-1)}.
\end{equation}
Thus, we may define
\begin{equation}\label{E29}
{\mathcal{L}^{-}}=\displaystyle(n-\zeta)\sqrt{\zeta}\Big[\frac{\partial}{\partial \zeta}-\left(-\frac{1}{2}+\frac{|m|}{2\zeta}
-\frac{n(|m|+\zeta)}{\zeta(n-\zeta)}\right)\Big]e^{-i\varphi}{, }
\end{equation}
with the following property
\begin{equation}\label{E30}
{\mathcal{L}^{-}}\psi_{nm}=\sqrt{n(n+|m|)(n+|m|-1)}\psi_{(n-1)(m-1)}.
\end{equation}
Accordingly we have found the ladder operators ${\mathcal{L}^{\pm}}$ which establish the connection between the wave functions $\psi_{nm}(\vec{\rho})$ and $\psi_{(n\pm 1)\, (m\pm 1)}(\vec{\rho})$.

\section{Concluding remarks}
In this work we have established the ladder operators for a charged particle moving in a two dimensional uniform magnetic field, but these operators depend on all spatial variables $\rho$ and $\varphi$. Finally we have reconsidered this quantum system from algebraic point of view by introducing a velocity operator.

\vskip 0.5cm
\noindent
{\large\bf Acknowledgments}: This work is partially supported by 20160978-SIP-IPN and 20170938-SIP-IPN, Mexico.

\end{document}